\documentclass[11pt,a4paper]{article}
\usepackage{jheppub}
\usepackage{tikz}

\usepackage{color}
\definecolor{math1}{rgb}{0.368417,0.506779,0.709798}
\definecolor{math2}{rgb}{0.880722,0.611041,0.142051}
\definecolor{math3}{rgb}{0.560181,0.691569,0.194885}
\definecolor{math4}{rgb}{0.922526,0.385626,0.209179}

\newcommand{\D}{\mathrm{d}}

\newcommand{\ket}[1]{\left| #1 \right\rangle}

\newcommand{\branket}[2]{\left\langle #1 \middle| #2 \right\rangle}

\renewcommand{\vec}{\boldsymbol}

\newcommand{\legerrorpoint}[3]{
\draw[color=#1,fill] (#2) circle (0.05cm) node [right,xshift=5] {\color{black}#3};
\draw[color=#1]      (#2) --+ (0, 0.2) --+ (0.05, 0.2) --+ (-0.05, 0.2);
\draw[color=#1]      (#2) --+ (0,-0.2) --+ (0.05,-0.2) --+ (-0.05,-0.2);
}

\newcommand{\MeV}{\mbox{MeV}}
\newcommand{\MS}{\overline{\rm MS}}
\newcommand{\Em}{E\hspace*{-6pt}/}

\begin{document}
\thispagestyle{empty}
\begin{flushright}
PSI-PR-16-14\\
ZU-TH 41/16\\
\today\\
\end{flushright}
\vspace{3em}
\begin{center}
{\Large\bf Fully differential NLO predictions for the rare muon decay}
\\
\vspace{3em}
{\sc G. M. Pruna$^a$, A.~Signer$^{a,b}$, Y. Ulrich$^{a,b}$
}\\[2em]
{\sl ${}^a$ Paul Scherrer Institut,\\
CH-5232 Villigen PSI, Switzerland \\
\vspace{0.3cm}
${}^b$ Physik-Institut, Universit\"at Z\"urich, \\
Winterthurerstrasse 190,
CH-8057 Z\"urich, Switzerland}
\setcounter{footnote}{0}
\end{center}
\vspace{2ex}
\begin{center}
\begin{minipage}[]{0.9\textwidth}
{} {\sc Abstract:} Using the automation program GoSam, fully
differential NLO corrections were obtained for the rare decay of the
muon $\mu\to e\nu\bar\nu ee$. This process is an important Standard
Model background to searches of the Mu3e collaboration for
lepton-flavour violation, as it becomes indistinguishable from the
signal $\mu\to 3e$ if the neutrinos carry little energy. With our NLO
program we are able to compute the branching ratio as well as
custom-tailored observables for the experiment. With minor
modifications, related decays of the tau can also be computed.
\end{minipage}
\end{center}

\setcounter{page}{1}


\bigskip

\section{Introduction}
The rare decay of the muon $\mu \to e \nu \nu e e$ is not one of those
Standard Model processes that has been in the limelight in the past
decades. The branching ratio $\mathcal{B} = (3.4\pm 0.4)\times
10^{-5}$ has been measured more than thirty years ago by the SINDRUM
collaboration~\cite{Bertl:1985mw, Olive:2016xmw} in the context of
searches for the lepton-flavour violating process $\mu \to 3 e$. For
the corresponding decay $\tau \to e \nu \nu e e$ a measurement
$\mathcal{B} = (2.8\pm 1.5)\times 10^{-5}$ is available from
CLEO~\cite{Alam:1995mt}, but for other leptonic five-body decays of
the $\tau$ only upper limits exist.

On the theory side, several tree-level calculations of the branching
ratios for the various rare decays have been made long
ago~\cite{Fishbane:1985xz, Kersch:1987dw, Dicus:1994dt}. These
calculations might have been sufficient for the previous experimental
situation. However, with the upcoming Mu3e~\cite{Blondel:2013ia,
  Berger:2014vba, Perrevoort:2016nuv} experiment that is dedicated to
the search for the lepton-flavour violating decay $\mu\to 3e$ at the
Paul Scherrer Institute (PSI) an improved theoretical description of
the rare muon decay is highly desirable for several reasons. First,
the rare muon decay is an interesting process in its own right as it
can be measured very precisely by Mu3e. Second, in the limit where the
neutrinos have very little energy, it is a background to the decay
$\mu\to 3e$.

In order to avoid infrared singularities and to get a reliable
prediction in all corners of phase space, it is necessary to keep finite
electron mass terms.  Furthermore, a realistic background study
benefits from a fully differential description of the rare muon
decay. This requires the computation of the corresponding matrix
elements and their implementation in a Monte Carlo program.

The squared matrix elements at tree level for the unpolarized rare
muon decay and a study of the energy spectrum have been published less
than ten years ago~\cite{Djilkibaev:2008jy}. For a more realistic
investigation of the background to $\mu\to 3e$, the polarization of
the muon has to be taken into account.  Even though the tree-level
calculation of the matrix elements in the Fermi theory is trivial,
these matrix elements have been made available to the Mu3e
collaboration only very recently~\cite{polmatel}. A tree-level study
of the rare muon and tau decay has been published at a later
stage~\cite{Flores-Tlalpa:2015vga}.

Typically the next-to-leading (NLO) order QED corrections are expected
to be at the percent level. However, in extreme regions of phase
space, as for example at the endpoint of the lepton energy spectrum,
corrections can be enhanced due to large logarithms. As seen in a
recent NLO calculation~\cite{Fael:2015gua} of the radiative muon and
tau decays $\ell\to \ell' \nu \nu \gamma$, QED corrections can easily
be as large as 10\%. Thus, a fully differential NLO Monte Carlo for
the rare decays of a polarized muon is an important ingredient to
fully exploit the future data to be taken by the Mu3e
collaboration. In this article we present such a calculation. We use
GoSam~\cite{Cullen:2014yla} to obtain the matrix elements and perform
the phase-space integration using FKS
subtraction~\cite{Frixione:1995ms, Frederix:2009yq}.

The paper is organized as follows: in section~\ref{sec:method} we will
discuss our methodology. Since this is a standard NLO calculation we
focus on changes to GoSam that were necessary to compute the required
one-loop matrix elements. In section~\ref{sec:ratios} we present some
branching ratios that were independently confirmed
by~\cite{BerneTBP}. Finally, in section~\ref{sec:distri} we present
some distributions and conclude in section~\ref{sec:conclusion}.

\section{Methodology}\label{sec:method}
We perform the computation in the Fierz-rearranged effective 4-Fermi
interaction, i.e. with the Lagrangian
\begin{align}
\mathcal{L} &
= \mathcal{L}_{\text{QED}} - \frac{4\, G_F}{\sqrt2} 
\left( \bar\psi_{\nu_\mu} \gamma^\mu P_L \psi_\mu \right)
\left( \bar\psi_e \gamma^\mu P_L \psi_{\nu_e} \right)\notag\\&
= \mathcal{L}_{\text{QED}} + \frac{4\, G_F}{\sqrt2} 
\left( \bar\psi_e \gamma^\mu P_L \psi_\mu \right)  
\left( \bar\psi_{\nu_\mu} \gamma^\mu P_L \psi_{\nu_e} \right)\, ,
\label{eq:fierzed}
\end{align}
where $P_L = (1-\gamma_5)/2$ is the usual left-handed projector. Even
though this is an effective theory, it can be shown that the Fermi
constant $G_F$ does not get renormalized and all QED corrections are
finite after the usual QED renormalization~\cite{BERMAN196220}.  The
QED Lagrangian $\mathcal{L}_{\text{QED}}$ includes electron $\psi_e$
and muon fields $\psi_\mu$ (and tau fields for the rare tau decay) but
no quark fields. The effect of the latter is very small for muon
decays. As mentioned in the introduction we keep $m_e \neq 0$.

\subsection{The rare decay}

At leading order the rare decay $\mu\to e\nu\bar\nu e e$ is given
through four diagrams. At NLO there are about 40 one-loop diagrams and
20 diagrams involving a real emission. These matrix elements are
generated using the automation tool GoSam~\cite{Cullen:2014yla} and
reduced at run time using \textsc{Ninja}~\cite{Mastrolia:2012bu,
  vanDeurzen:2013saa, Peraro:2014cba} or
\texttt{golem95}~\cite{Binoth:2008uq, Mastrolia:2010nb}. Scalar
integrals were computed using OneLOop~\cite{vanHameren:2009dr,
  vanHameren:2010cp}.

The arising soft singularities from the real emission diagrams are
treated using FKS subtraction~\cite{Frixione:1995ms,
  Frederix:2009yq}. In the absence of collinear singularities, the FKS
method is particularly simple as it generally treats soft and
collinear singularities separately.

Finally, a numerical integration of the full phase space is carried
out using VEGAS~\cite{Lepage:1980jk}. This allows for the production
of any differential observables with arbitrary cuts.  Instead of a
general purpose phase-space generator that creates momenta
recursively, we use a tailored phase-space generator. It is designed
such that pseudo singularities, which cause numerical instabilities
due to the smallness of $m_e$, are aligned with the variables of the
VEGAS integrator to optimally utilize the VEGAS adaption.

For the renormalization of the lepton masses we always choose the
on-shell scheme. The QED coupling can be renormalized either in the
on-shell scheme $\alpha = \alpha_{\rm os}$ or in the 
$\MS$ scheme $\bar\alpha =\alpha_{\MS}(\mu=m_\mu)$. The results can
of course easily be converted using
\begin{align}
\bar\alpha &= \alpha \left(1 +
\frac{\alpha}{3\pi}  \log\frac{m_\mu^2}{m_e^2} \right).
\end{align}
Since GoSam returns the NLO matrix elements in the four-dimensional
helicity scheme (FDH)~\cite{Bern:1991aq}, care has to be taken to use
this scheme also for the external wave-function renormalization and
the real corrections~\cite{Signer:2008va}.

Regarding the $\gamma_5$ that is present in \eqref{eq:fierzed} we note
that the radiative corrections to the axial vector contribution are
related to those of the vector contribution by setting $m_e \to
-m_e$~\cite{BERMAN196220}. This is due to the fact that the
simultaneous transformation $\psi_e\to \gamma_5 \psi_e$ and
$m_e\to-m_e$ leaves $\mathcal{L}_{\text{QED}}$ invariant but exchanges
the vector and axial-vector currents.  Using an anticommuting
$\gamma_5$, as done in GoSam~\cite{Cullen:2010jv}, respects this
relation and hence avoids problems with $\gamma_5$ in this particular
case.

\subsection{Changes in GoSam}
The GoSam package is designed to compute one-loop amplitudes for
processes involving an arbitrary number of particles in QED and
QCD. To use the 4-Fermi interaction~\eqref{eq:fierzed}, the
corresponding Feynman rules have to be incorporated into GoSam's model
file.

GoSam applies the spinor helicity formalism using light-cone
decomposition. For a lepton of mass $m$ the momentum $q^\rho$ is decomposed
with the help of two light-like vectors $\ell^\rho$ and $n^\rho$ as
\begin{align}
q^\rho &= \ell^\rho +\frac{m^2}{2 (q\cdot n)}  \, {n^\rho}
= \ell^\rho +\frac{m^2}{2 (\ell\cdot n)}  \, {n^\rho}  \, .
\end{align}
Then the spinor for the massive lepton is written in terms of the
usual massless spinors as 
\begin{align}
u_\pm(q) &= \ket{\ell^\pm} +\frac{m}{\branket{\ell^\pm}{n^\mp}} \ket{n^\mp}\,.
\end{align}
Unfortunately, the reference vector $n^\rho$ relative to which the
polarization is defined is always chosen by GoSam to be an internal
vector, typically the momentum of a massless particle in the
process. While this greatly simplifies the amplitudes that are
evaluated, it is inconvenient to compute polarized observables this
way. An external vector as the reference vector is needed. This can be
changed by adapting the GoSam process template such that the
light-cone decomposition of massive fermions is done with respect to
the new external light-like reference instead of an internal
light-like vector.

Additionally, the actual GoSam code needs to be adapted to allow for
massive lepton counterterms. 

\section{Branching ratios}\label{sec:ratios}

In this section we present some results for the branching ratio
\begin{align}
\mathcal{B} 
&= \Gamma(\mu^+\to e^+\bar\nu_\mu\nu_e e^+ e^- )/\Gamma_\mu\, ,
\end{align}
where $\Gamma_\mu=1/\tau_\mu$ with $\tau_\mu=2.19698\times 10^{-6}$~s
is the experimentally measured width of the muon.  We have used the
standard values for the various inputs~\cite{Olive:2016xmw}:
\begin{equation}
\begin{aligned}
m_e &= 0.51099893~\MeV, & m_\mu &= 105.65837~\MeV, \\
G_F &= 1.166379\times 10^{-11}~\MeV^{-2}, & 
\alpha &= 1/137.0356 \, .
\end{aligned}
\end{equation}

In addition to the full branching ratio we also consider the branching
ratio with a constraint on the invisible energy
\begin{align}
\Em \equiv m_\mu-E_1-E_2-E_3\, ,
\end{align}
where $E_1 \ge E_2$ are the energies of the positrons and $E_3$ is the
energy of the electron, respectively.

In table~\ref{tab:branching} we summarize the branching ratios in the
on-shell scheme for various cuts on $\Em$. The error indicated in the
table is the numerical error of the Monte Carlo integration only. For
ever more stringent cuts on $\Em$ it becomes increasingly more
difficult to obtain stable numerical results with our general purpose
Monte Carlo program.  As can be seen, the corrections are moderate if
no cut is applied, but become substantial for stringent cuts on
$\Em$. Results for the branching ratio with such stringent cuts on
$\Em$ are particularly relevant for a background estimate of the Mu3e
experiment.  Furthermore, these results allow for a comparison
with~\cite{Djilkibaev:2008jy} at leading order and
with~\cite{BerneTBP} at NLO. We find perfect agreement in all cases,
assuming that \cite{Djilkibaev:2008jy} used $\Gamma_\mu^{(0)} = G_F^2
m_\mu^5/(192 \pi^3)$ to normalize the branching ratio.

\begin{figure}
\centering
\begin{tabular}{c|c|c|c||c}
$\mathcal{B}$ & LO & \multicolumn{2}{|c||}{NLO correction} & Correction \cite{BerneTBP} \\\hline
no cuts                           & {$3.605(1) \cdot 10^{-5}  $}{} & {$-6.74(4) \cdot 10^{-8} $}{} & {$-0.19$}{\%} & {$-6.69(5) \cdot 10^{-8} $}{} \\
$\Em < 100\cdot m_e$ & {$2.121(1) \cdot 10^{-6} $}{} & {$-9.46(2) \cdot 10^{-8} $}{} &  {$-4.5$}{\%} & {$-9.47(6) \cdot 10^{-9} $}{} \\
$\Em < 50\cdot m_e$  & {$7.153(2) \cdot 10^{-9} $ }{} & {$-4.57(1) \cdot 10^{-10}$}{} &  {$-6.4$}{\%} & {$-4.55(3) \cdot 10^{-10}$}{} \\
$\Em < 20\cdot m_e$  & {$2.119(1) \cdot 10^{-11}$}{} & {$-2.14(4) \cdot 10^{-12}$}{} & {$-10.1$}{\%} & {$-2.17(1) \cdot 10^{-12}$}{} \\
$\Em < 10\cdot m_e$  & {$3.071(1) \cdot 10^{-13}$}{} & {$-4.05(1) \cdot 10^{-14}$}{} & {$-13.2$}{\%} & {$-4.04(2) \cdot 10^{-14}$}{} \\\hline
\end{tabular}
\renewcommand{\figurename}{Table}
\caption{The branching ratio of the rare decay in the on-shell scheme
  for various cuts on the invisible energy $\Em$. }
\label{tab:branching}

\end{figure}

Of course, we can also compute the leptonic five-body branching ratios
of the $\tau$. As an example we give 
\begin{align}
\label{eq:BR2}
\mathcal{B}_{\tau e} 
&= \Gamma(\tau^+\to e^+ \bar\nu_\tau\nu_e e^+ e^-)/\Gamma_\tau\, 
= \big(4.249(1) - 0.004(1)\big)\times 10^{-5}\, .
\end{align}
The first and second terms are the LO and NLO contributions in the
on-shell scheme, respectively. Again, the errors are numerical errors
only and we have used the experimental value for
$\Gamma_\tau=1/\tau_\tau$ with $\tau_\tau = 2.903\times 10^{-13}$~s
and $m_\tau = 1776.8~\MeV$.  As for the muon case, in the on-shell
scheme the corrections are very small in the absence of additional
cuts. In the $\MS$-scheme, the NLO corrections are about $-3.5\%$. We
note that the muonic and hadronic photon vacuum polarization
contributions both give a very small effect $\mathcal{O}(10^{-9})$ to
$\mathcal{B}_{\tau e}$.  The latter have been estimated using the
Fortran code {\tt hadr5n12}~\cite{Jegerlehner:1985gq,
  Burkhardt:1989ky, Jegerlehner:2006ju}.  The result given in
\eqref{eq:BR2} is to be compared with the experimental result from
CLEO $\mathcal{B}_{\tau e} = (2.8\pm 1.5)\times
10^{-5}$~\cite{Alam:1995mt}. For other leptonic decays of the $\tau$
only upper limits are available from experiment. They can be computed
easily with our code even if processes with two different lepton
flavours in the final state require trivial modifications.

\section{Distributions}\label{sec:distri}
In addition to branching ratios, differential distributions with
arbitrary cuts can also be computed with our Monte Carlo program. To
provide some examples, the following distributions will be considered:
\begin{itemize}
\item $\D\mathcal{B}/\D \Em$: The invisible energy
  distribution is important to correctly estimate the background from
  the rare decay to the LFV decay $\mu^+ \to e^+  e^+  e^-$.
\item $\D\mathcal{B}/\D x_i$ where $x_i \equiv 2 E_i/m_\mu$: The
  momentum fraction distributions of the three charged final-state
  leptons can be used to experimentally discriminate between different
  LFV models.
\item $\D\mathcal{B}/\D \cos\theta_i$ where $\theta_i \equiv
  \sphericalangle(\vec q_i, \hat{\vec z})$ is the angle between the
  $z$-axis and the momentum $q_i$ of the various final-state leptons:
  The angular distributions of the charged leptons can be used to
  study the $V-A$ structure of the Fermi interaction, or in the case
  of LFV, the Lorentz structure of the effective operator.
\end{itemize}
For illustration, these calculations were carried out using the
polarization  $\vec s={-0.85}\,\hat{\vec z}$ and imposing
\begin{align}
\label{eq:cuts}
E_i > {10~\MeV} \quad\text{and}\quad|\cos\theta_i| < 0.8\,.
\end{align}
These cuts and the chosen polarization are meant to simulate the
expected situation for the Mu3e experiment. While the results in
section~\ref{sec:ratios} were independent of the polarization, the
angular distributions will of course be affected.

\subsection{Invisible energy spectrum}

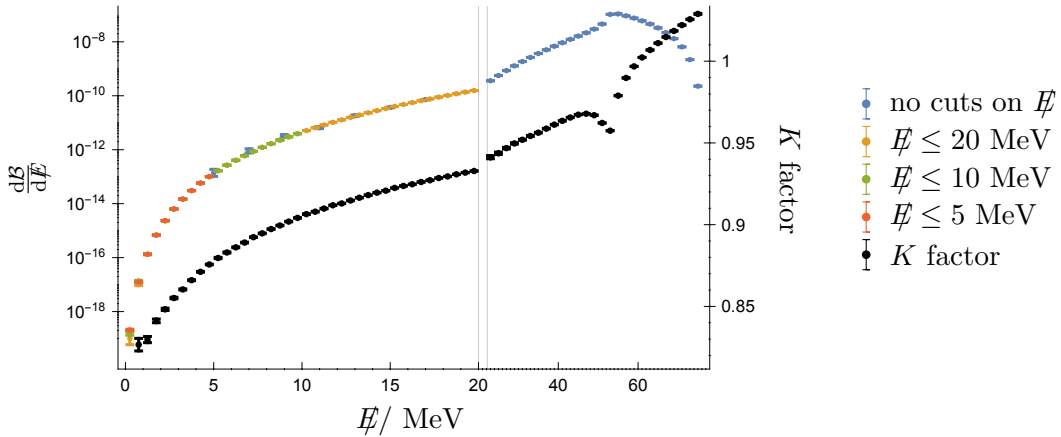
\begin{figure}[t]
\centering

\begin{tikzpicture}
\node [anchor=south] at (0,0) {\includegraphics[width=0.6\textwidth]{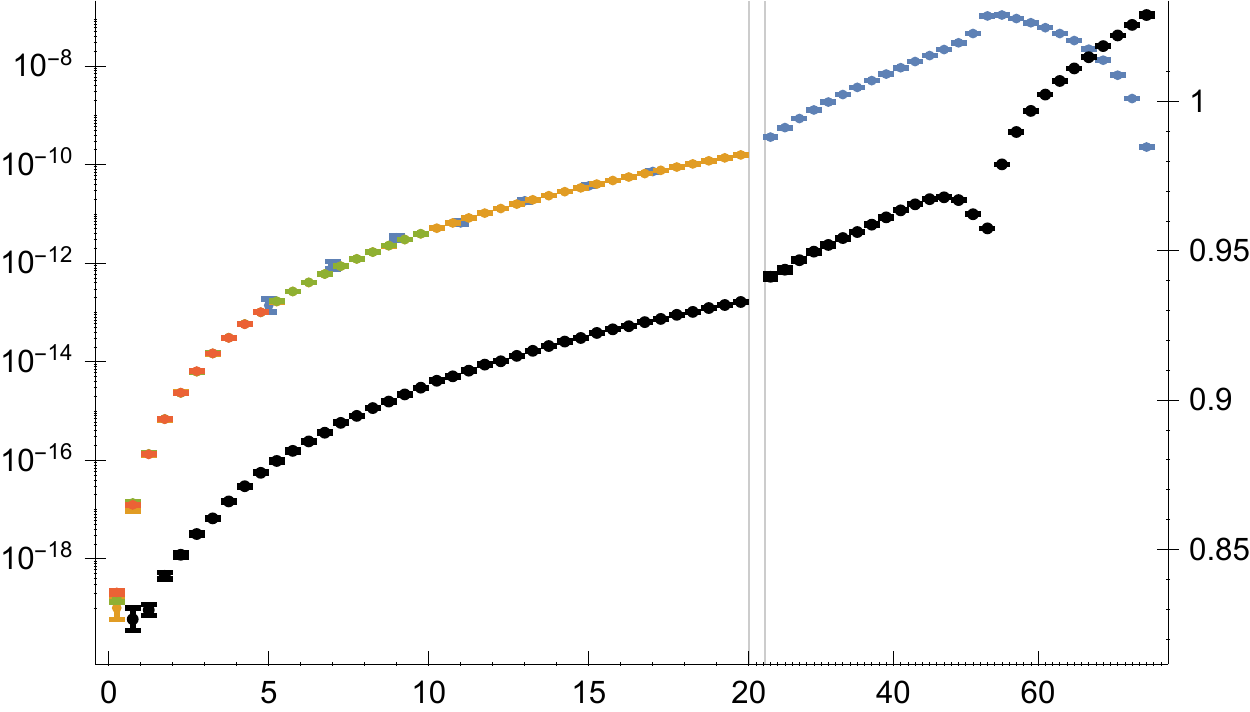}};
\node at (0,-0.2) {$\Em / {~\MeV}$};
\node[rotate= 90] at (-5,3) {$\frac{\D\mathcal{B}}{\D \Em}$};
\node[rotate=-90] at ( 5,3) {$K$ factor};

\legerrorpoint{math1}{6,4.0}{no cuts on $\Em$}
\legerrorpoint{math2}{6,3.5}{$\Em \le {20~\MeV}$}
\legerrorpoint{math3}{6,3.0}{$\Em \le {10~\MeV}$}
\legerrorpoint{math4}{6,2.5}{$\Em \le {5~\MeV}$}
\legerrorpoint{black}{6,2.0}{$K$ factor}
\end{tikzpicture}

\caption{The differential decay distribution w.r.t. the invisible
  $\Em$ at NLO in blue, orange, green and red (see text) and the $K$
  factor NLO/LO in black. To emphasize the low energy tail, the
  scaling is broken at $\Em = {20~\MeV}$. The error bars indicate the
  numerical error of the Monte Carlo integration.}
\label{fig:mu3e:invis}
\end{figure}

One of the most important distributions, $\D\mathcal{B}/\D \Em$, is
shown in figure~\ref{fig:mu3e:invis} with numerical errors indicated
by the error bars.  This distribution falls sharply in the region
$\Em\to 0$ that is particularly important for the Mu3e experiment.
Thus, using a standard configuration (shown in blue in
figure~\ref{fig:mu3e:invis}) it is challenging to get enough
statistics in the low energy tail to obtain predictions with
reasonable statistical uncertainties. To improve the tail, we focused
on the low energy region by imposing an additional cut on $\Em$ of
${20~\MeV}$ (shown in orange), ${10~\MeV}$ (shown in green) and
${5~\MeV}$ (shown in red) and combined the results.

As can be seen from figure~\ref{fig:mu3e:invis}, the NLO corrections
are negative except for a small region of maximal $\Em$. In the
low-energy tail, the corrections exceed $-10\%$.  Hence, there are
substantially fewer background events to $\mu\to 3e$ from the rare
decay than expected from tree-level simulations.  The cuts $E_i >
{10~\MeV}$ are the reason for the sharp fall of the distribution at
$\Em\to m_\mu - 30~\MeV$. The kink in the distribution is at about
$m_\mu/2$, shifted to somewhat lower values due to the effects of the
non-vanishing electron mass. In fact, due to the additional real
radiation of a photon, the NLO corrections amount to shifting the
distribution $\D\mathcal{B}/\D \Em$ to higher energies.

\begin{figure}[ht]

\scalebox{0.72}{
\begin{tikzpicture}
\node [anchor=south] at (0,0) {\includegraphics[width=258pt]{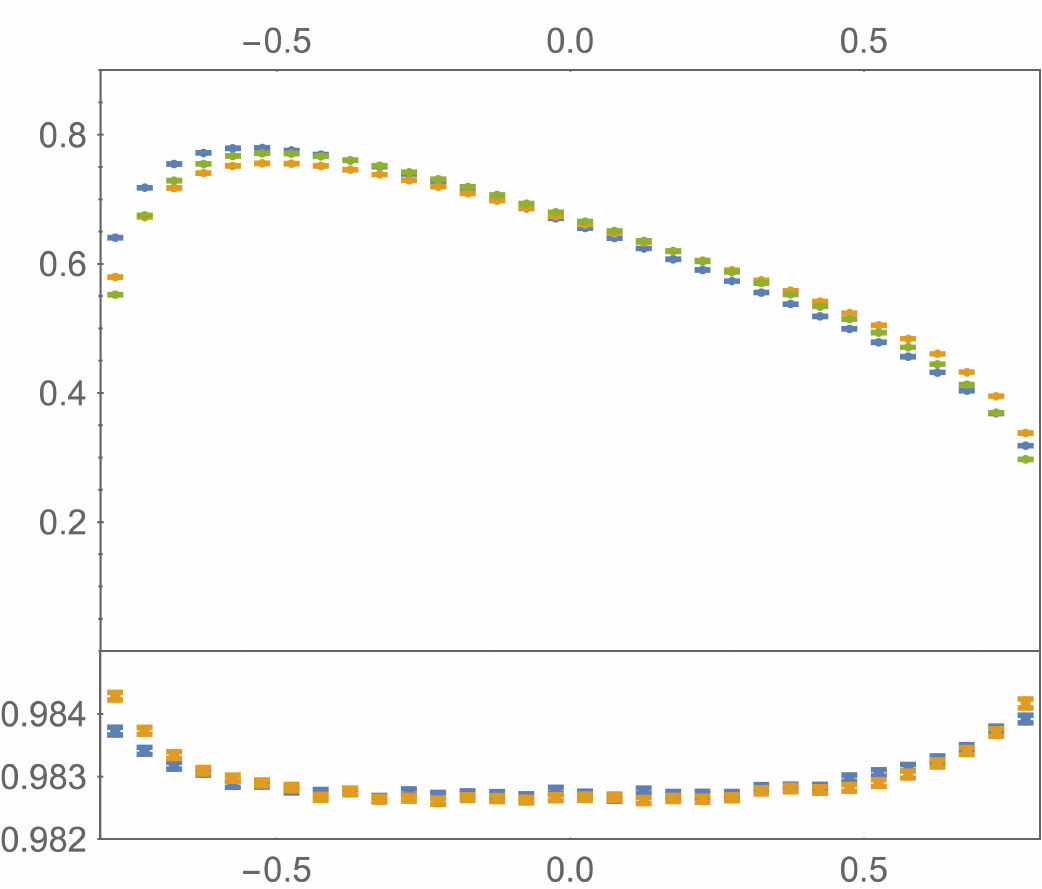} 
             \hspace*{2cm}     \includegraphics[width=258pt]{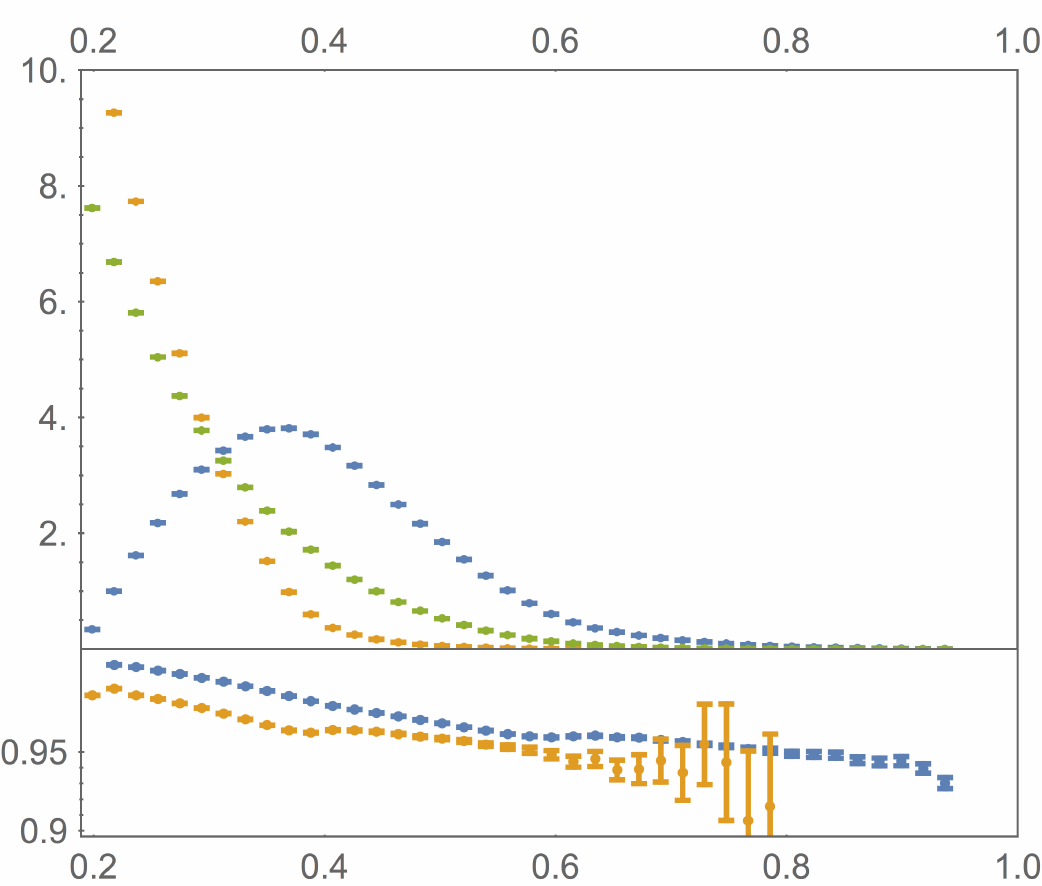}};
\node at (0,-0.25) {No cut on $\Em$ };
\node[rotate= 90] at (-10.5,4) {\Large$\frac{1}{\mathcal{B}}\frac{\D\mathcal{B}}{\D\cos\theta_i}$};
\node[rotate= 90] at (0.7,4) {\Large$\frac{1}{\mathcal{B}}\frac{\D\mathcal{B}}{\D x_i}$};
\legerrorpoint{math1}{-2.5,8}{hard $e^+$}
\legerrorpoint{math2}{ 0.0,8}{soft $e^+$}
\legerrorpoint{math3}{ 2.5,8}{$e^-$}
\end{tikzpicture}
}

\scalebox{0.72}{
\begin{tikzpicture}
\node [anchor=south] at (0,0) {\includegraphics[width=258pt]{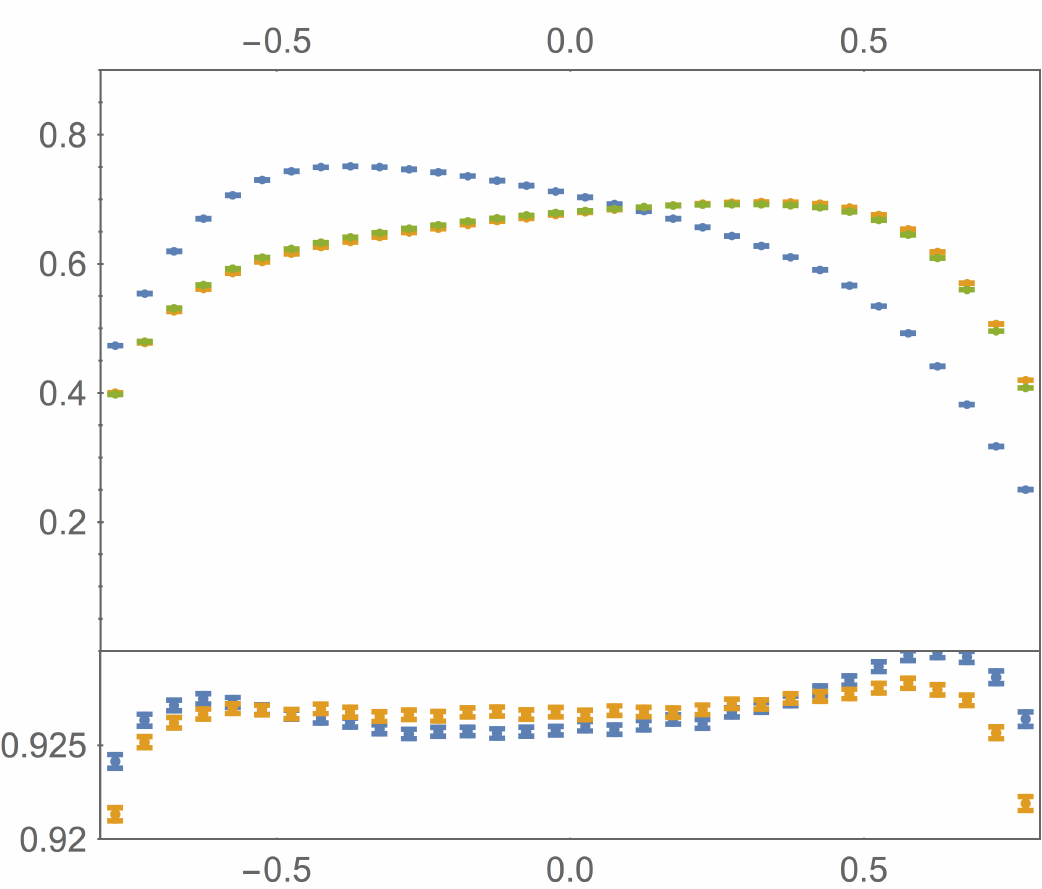} 
             \hspace*{2cm}     \includegraphics[width=258pt]{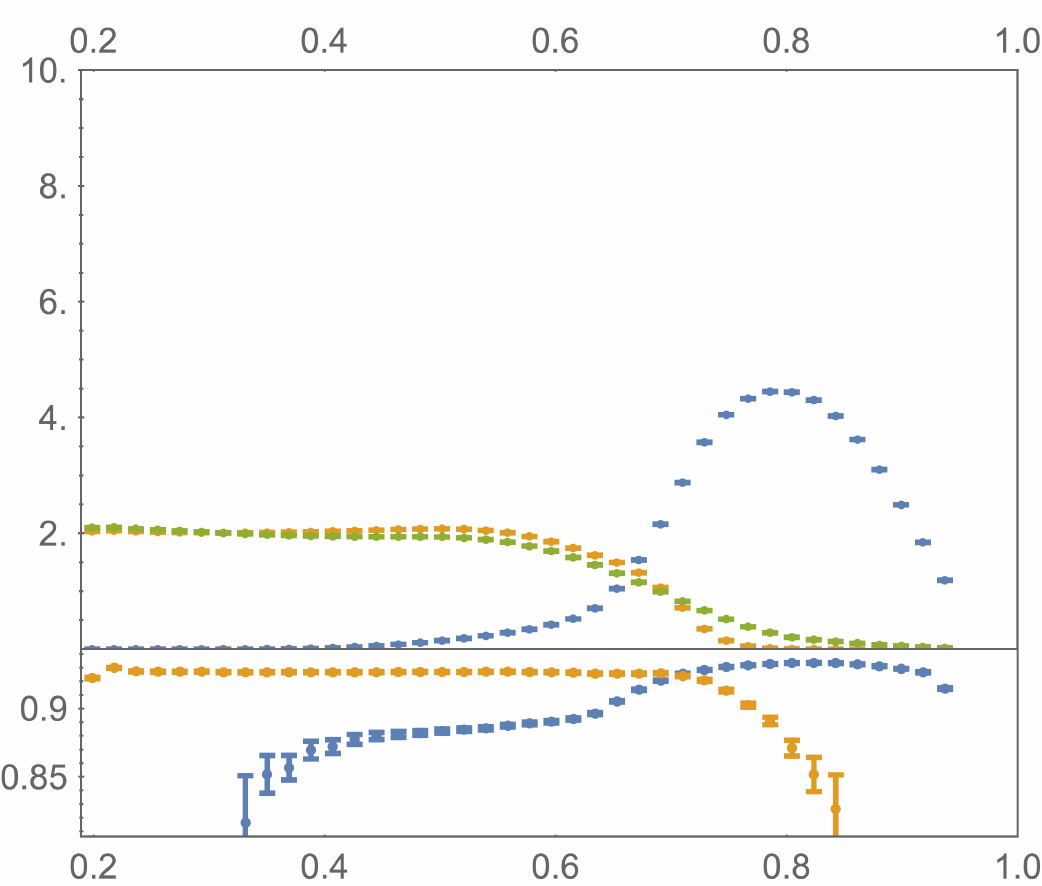}};
\node at (0,-0.25) {$\Em \le {20~\MeV}$};
\node[rotate= 90] at (-10.5,4) {\Large$\frac{1}{\mathcal{B}}\frac{\D\mathcal{B}}{\D\cos\theta_i}$};
\node[rotate= 90] at (0.7,4) {\Large$\frac{1}{\mathcal{B}}\frac{\D\mathcal{B}}{\D x_i}$};
\end{tikzpicture}
}

\scalebox{0.72}{
\begin{tikzpicture}
\node [anchor=south] at (0,0) {\includegraphics[width=258pt]{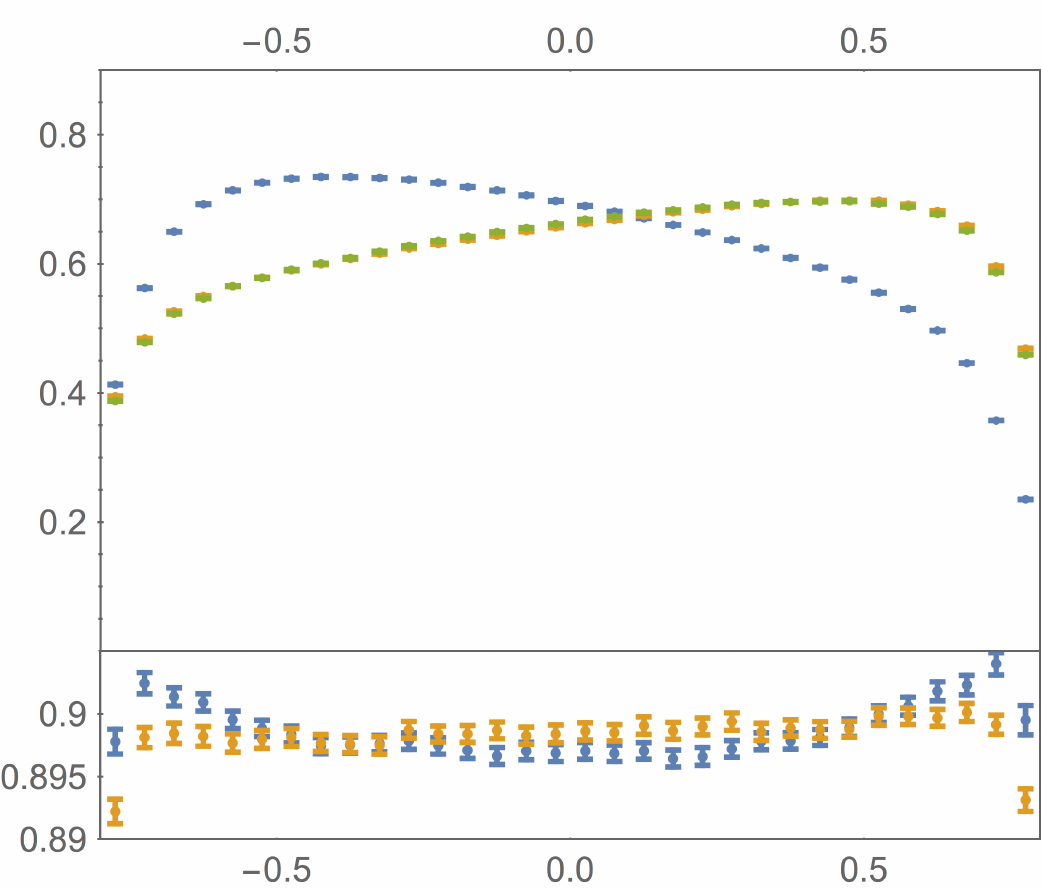} 
             \hspace*{2cm}     \includegraphics[width=258pt]{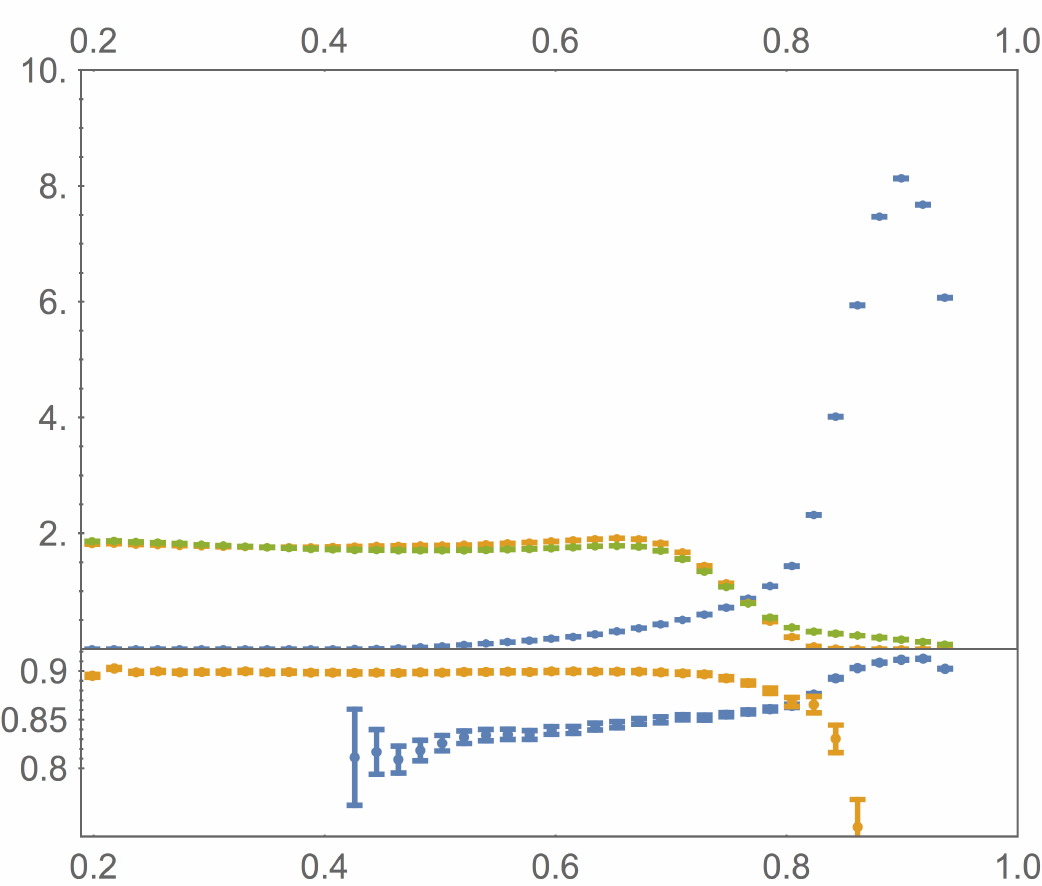}};
\node at (0,-0.25) {$\Em \le {10~\MeV}$};
\node at (-5.2,-0.25) {$\cos\theta_i$};
\node at (6,-0.25) {$x_i = 2E_i/m_\mu$};
\node[rotate= 90] at (-10.5,4) {\Large$\frac{1}{\mathcal{B}}\frac{\D\mathcal{B}}{\D\cos\theta_i}$};
\node[rotate= 90] at (0.7,4) {\Large$\frac{1}{\mathcal{B}}\frac{\D\mathcal{B}}{\D x_i}$};
\end{tikzpicture}
}

\caption{Angular spectrum (left panel) and energy spectrum (right
  panel) of the final state leptons at NLO (blue: more energetic
  positron; orange: less energetic positron; green: electron). In the
  top panel no cut on $\Em$ is imposed whereas in the middle and lower
  panel a cut of $\Em \le {20~\MeV}$ and $\Em \le {10~\MeV}$ is
  applied, respectively. The $K$-factors for the positrons are shown
  in the subpanels.  The error bars indicate the
  numerical error of the Monte Carlo integration.}\label{fig:mu3e:dist}

\end{figure}
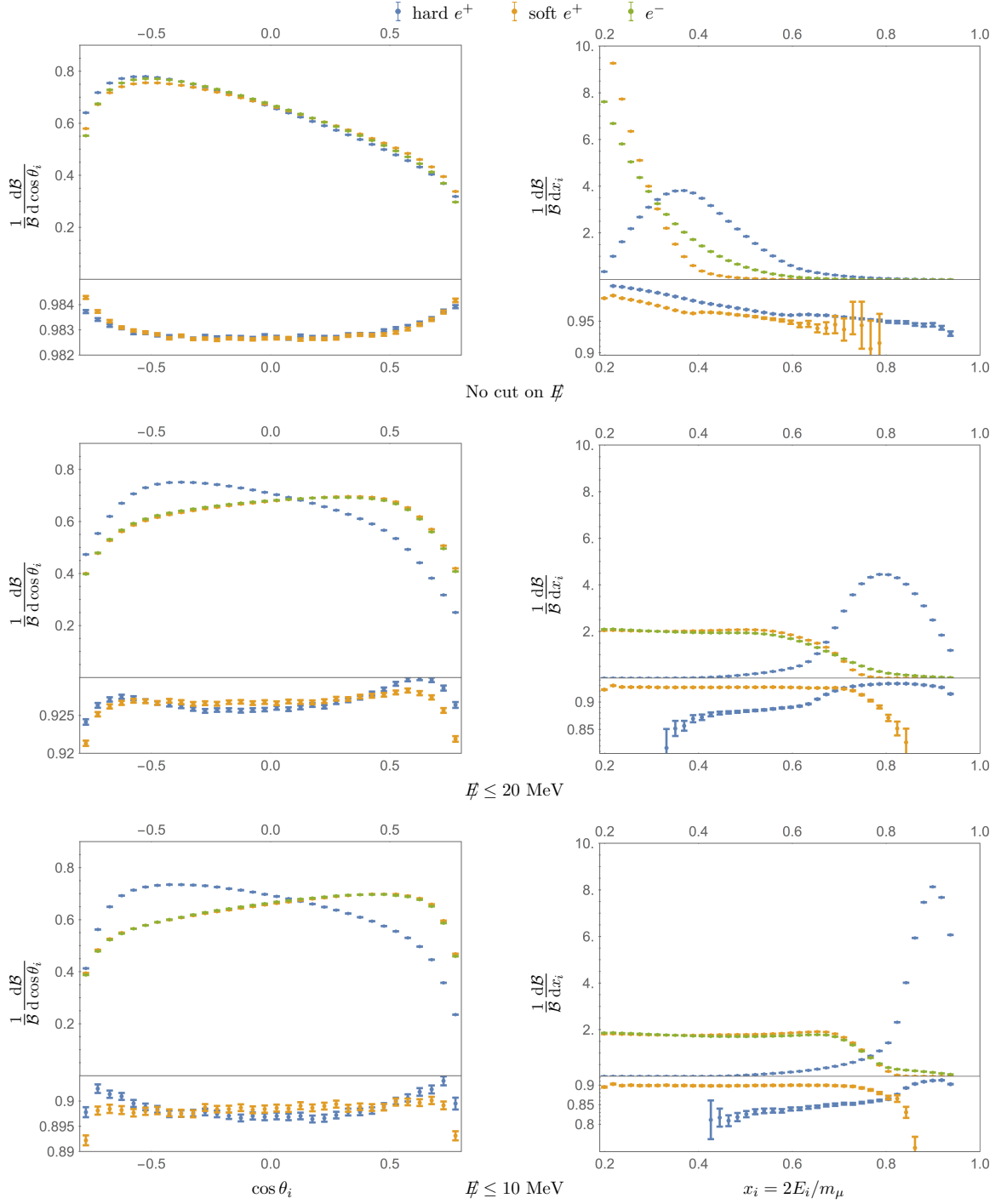

\subsection{Momentum fraction and angular distribution}

As an example of a distribution where the polarization of the muon has
an effect, we consider the angular distribution of the charged
leptons. We consider three cases: no cut on the missing energy, a cut
$\Em \le {20~\MeV}$ and finally a cut $\Em \le {10~\MeV}$, focusing
on events in the tail of the $\Em$ distribution. In the left panels of
figure~\ref{fig:mu3e:dist} we show the normalized distributions
$\D\mathcal{B}/{\D\cos\theta}$ at NLO for the more energetic positron
(blue), the less energetic positron (orange) and the electron (green).
In the right panel of figure~\ref{fig:mu3e:dist} we show the
normalized energy fraction distributions. The $K$-factor for the
positrons is shown in the subpanels. The electron's $K$-factor is
virtually indistinguishable to the one of the soft positron.

As observed already in section~\ref{sec:ratios}, the size of the
corrections tends to increase if cuts are applied.  With the cuts of
\eqref{eq:cuts} the corrections for the full branching ratio amount to
$-1.7\%$ ($-3.4\%$) in the on-shell scheme ($\MS$ scheme). If in
addition we require $\Em \le {20~\MeV}$ the corrections increase to
$-7.3\%$ ($-9.1\%$) and for $\Em \le {10~\MeV}$ they are $-10.2\%$
($-11.9\%$). At NLO, the difference between results in the on-shell
and $\MS$ scheme is less than $0.5\%$.

With the chosen polarization of the muon, if no cut on $\Em$ is
applied, all final state leptons prefer to be emitted backwards, as is
the case for the normal muon decay.  However, with a cut on $\Em$ the
hard positron behaves distinctly different to the other two
leptons. Loosely speaking, this can be understood by noting that the
more energetic positron is the primary positron, whereas the soft
positron (and the electron) are produced from the conversion of the
internal photon. As the cut on $\Em$ becomes more restrictive, this
effect becomes more pronounced. The soft positron behaves ever more
similar to the electron. This can also be seen in the energy
distribution where the hard positron behaves drastically different
from the soft positron and the electron, as shown in the right panel
of figure~\ref{fig:mu3e:dist}.  Furthermore, the size of the NLO
corrections increases with more restrictive cuts, in analogy to what
can be seen in table~\ref{tab:branching} and
figure~\ref{fig:mu3e:invis}.

We stress that the distributions shown in figures~\ref{fig:mu3e:invis}
and \ref{fig:mu3e:dist} just serve as an illustration and that we can
compute any distribution with arbitrary cuts at NLO. In regions of
phase space, where there are not many events, the numerical error is
large and a dedicated run is needed to improve the numerical
precision.

\section{Conclusion}\label{sec:conclusion}
We have presented a Monte Carlo code for the rare polarized muon decay
$\mu\to e \nu\bar\nu ee$. This code allows to compute branching ratios
and distributions with arbitrary cuts at NLO in the Fermi theory.  Our
results for the branching ratio have been compared with
Ref.~\cite{BerneTBP} and full agreement has been found.

If no stringent cuts are applied to the final state, the size of the
corrections in the on-shell scheme is very modest, i.e. at the order
of $1\%$. Whilst it is notoriously difficult to give a reliable
theoretical error of higher-order calculations, this is a good
indication that the NLO calculation in this case provides a
theoretical prediction with an error well below $1\%$. Corrections
beyond NLO in the Fermi theory are very unlikely to be larger. Effects
beyond the Fermi theory are suppressed by the large electroweak mass
scale $M_{\rm ew}$ and are of the order $(m_\mu/M_{\rm ew})^2\sim
10^{-6}$.

In connection with background studies for the Mu3e experiment, it is
often important to consider stringent cuts. In particular, the region
of phase space where the invisible energy $\Em$ is very small is of
importance. With such cuts the corrections can easily reach $10\%$ or
even more. The fact that the corrections are large and negative is
favourable to $\mu\to 3e$ searches. However, the size of the
corrections indicates that in this case corrections beyond NLO in the
Fermi theory are likely in the region of $1\%$.

From a phenomenological point of view five-body leptonic decays of the
muon are much more important than the corresponding decays of the
tau. Nevertheless, the latter can also be obtained at NLO with minor
modifications of our code.

\acknowledgments{ We are grateful to Gionata Luisoni for his
  invaluable help while modifying GoSam and to Angela~Papa and
  Giada~Rutar for very useful discussions regarding the Mu3e
  experiment. We would further like to thank Matteo Fael and Christoph
  Greub for sharing their results for comparison and validation as
  well as helpful discussions. GMP and YU are supported by the Swiss
  National Science Foundation (SNF) under contract 200021\_160156 and
  200021\_163466, respectively.}

\bibliographystyle{JHEP}
\bibliography{references}{}

\providecommand{\href}[2]{#2}\begingroup\raggedright\begin{thebibliography}{10}

\bibitem{Bertl:1985mw}
{\scshape SINDRUM} collaboration, W.~H. Bertl et~al., \emph{{Search for the
  Decay $\mu^+ \to e^+ e^+ e^-$}},
  \href{http://dx.doi.org/10.1016/0550-3213(85)90308-6}{\emph{Nucl. Phys.} {\bf
  B260} (1985) 1--31}.

\bibitem{Olive:2016xmw}
C.~Patrignani, \emph{{Review of Particle Physics}},
  \href{http://dx.doi.org/10.1088/1674-1137/40/10/100001}{\emph{Chin. Phys.}
  {\bf C40} (2016) 100001}.

\bibitem{Alam:1995mt}
{\scshape CLEO} collaboration, M.~S. Alam et~al., \emph{{Tau decays into three
  charged leptons and two neutrinos}},
  \href{http://dx.doi.org/10.1103/PhysRevLett.76.2637}{\emph{Phys. Rev. Lett.}
  {\bf 76} (1996) 2637--2641}.

\bibitem{Fishbane:1985xz}
P.~M. Fishbane and K.~J.~F. Gaemers, \emph{{Calculation of the Decay $\mu^- \to
  e^- e^+ e^- \nu_\mu \bar{\nu}_e$}},
  \href{http://dx.doi.org/10.1103/PhysRevD.33.159}{\emph{Phys. Rev.} {\bf D33}
  (1986) 159}.

\bibitem{Kersch:1987dw}
A.~Kersch, N.~Kraus and R.~Engfer, \emph{{Analysis of the Rare Allowed Muon
  Decay $\mu^+ \to e^+ \nu_e \bar{\nu}_\mu e^+ e^-$}},
  \href{http://dx.doi.org/10.1016/0375-9474(88)90556-8}{\emph{Nucl. Phys.} {\bf
  A485} (1988) 606--620}.

\bibitem{Dicus:1994dt}
D.~A. Dicus and R.~Vega, \emph{{Standard Model decays of tau into three charged
  leptons}}, \href{http://dx.doi.org/10.1016/0370-2693(94)91389-7}{\emph{Phys.
  Lett.} {\bf B338} (1994) 341--348},
  [\href{http://arxiv.org/abs/hep-ph/9402262}{{\tt hep-ph/9402262}}].

\bibitem{Blondel:2013ia}
A.~Blondel, A.~Bravar, M.~Pohl, S.~Bachmann, N.~Berger et~al., \emph{{Research
  Proposal for an Experiment to Search for the Decay $\mu \to eee$}},
  \href{http://arxiv.org/abs/1301.6113}{{\tt 1301.6113}}.

\bibitem{Berger:2014vba}
{\scshape Mu3e} collaboration, N.~Berger, \emph{{The Mu3e Experiment}},
  \href{http://dx.doi.org/10.1016/j.nuclphysbps.2014.02.007}{\emph{Nucl.Phys.Proc.Suppl.}
  {\bf 248-250} (2014) 35--40}.

\bibitem{Perrevoort:2016nuv}
{\scshape Mu3e} collaboration, A.-K. Perrevoort, \emph{{Status of the Mu3e
  Experiment at PSI}},
  \href{http://dx.doi.org/10.1051/epjconf/201611801028}{\emph{EPJ Web Conf.}
  {\bf 118} (2016) 01028}, [\href{http://arxiv.org/abs/1605.02906}{{\tt
  1605.02906}}].

\bibitem{Djilkibaev:2008jy}
R.~M. Djilkibaev and R.~V. Konoplich, \emph{{Rare Muon Decay $\mu^+ \to e^+ e^-
  e^+ \nu_e \bar\nu_\mu$}},
  \href{http://dx.doi.org/10.1103/PhysRevD.79.073004}{\emph{Phys.Rev.} {\bf
  D79} (2009) 073004}, [\href{http://arxiv.org/abs/0812.1355}{{\tt
  0812.1355}}].

\bibitem{polmatel}
G.~M. Pruna and A.~Signer, {\emph{unpublished} (2015) }.

\bibitem{Flores-Tlalpa:2015vga}
A.~Flores-Tlalpa, G.~Lopez~Castro and P.~Roig, \emph{{Five-body leptonic decays
  of muon and tau leptons}},
  \href{http://dx.doi.org/10.1007/JHEP04(2016)185}{\emph{JHEP} {\bf 04} (2016)
  185}, [\href{http://arxiv.org/abs/1508.01822}{{\tt 1508.01822}}].

\bibitem{Fael:2015gua}
M.~Fael, L.~Mercolli and M.~Passera, \emph{{Radiative $\mu$ and $\tau$ leptonic
  decays at NLO}}, \href{http://dx.doi.org/10.1007/JHEP07(2015)153}{\emph{JHEP}
  {\bf 07} (2015) 153}, [\href{http://arxiv.org/abs/1506.03416}{{\tt
  1506.03416}}].

\bibitem{Cullen:2014yla}
G.~Cullen, H.~van Deurzen, N.~Greiner, G.~Heinrich, G.~Luisoni et~al.,
  \emph{{GoSam-2.0: a tool for automated one-loop calculations within the
  Standard Model and beyond}},
  \href{http://dx.doi.org/10.1140/epjc/s10052-014-3001-5}{\emph{Eur.Phys.J.}
  {\bf C74} (2014) 3001}, [\href{http://arxiv.org/abs/1404.7096}{{\tt
  1404.7096}}].

\bibitem{Frixione:1995ms}
S.~Frixione, Z.~Kunszt and A.~Signer, \emph{{Three jet cross-sections to
  next-to-leading order}},
  \href{http://dx.doi.org/10.1016/0550-3213(96)00110-1}{\emph{Nucl.Phys.} {\bf
  B467} (1996) 399--442}, [\href{http://arxiv.org/abs/hep-ph/9512328}{{\tt
  hep-ph/9512328}}].

\bibitem{Frederix:2009yq}
R.~Frederix, S.~Frixione, F.~Maltoni and T.~Stelzer, \emph{{Automation of
  next-to-leading order computations in QCD: The FKS subtraction}},
  \href{http://dx.doi.org/10.1088/1126-6708/2009/10/003}{\emph{JHEP} {\bf 10}
  (2009) 003}, [\href{http://arxiv.org/abs/0908.4272}{{\tt 0908.4272}}].

\bibitem{BerneTBP}
M.~{Fael} and C.~{Greub}, \emph{Next-to-leading order prediction for the decay
  $\mu\to e\, (e^+ e^-) \, \nu \bar\nu$}.

\bibitem{BERMAN196220}
S.~Berman and A.~Sirlin, \emph{Some considerations on the radiative corrections
  to muon and neutron decay},
  \href{http://dx.doi.org/http://dx.doi.org/10.1016/0003-4916(62)90114-8}{\emph{Annals
  of Physics} {\bf 20} (1962) 20 -- 43}.

\bibitem{Mastrolia:2012bu}
P.~Mastrolia, E.~Mirabella and T.~Peraro, \emph{{Integrand reduction of
  one-loop scattering amplitudes through Laurent series expansion}},
  \href{http://dx.doi.org/10.1007/JHEP11(2012)128,
  10.1007/JHEP06(2012)095}{\emph{JHEP} {\bf 06} (2012) 095},
  [\href{http://arxiv.org/abs/1203.0291}{{\tt 1203.0291}}].

\bibitem{Peraro:2014cba}
T.~Peraro, \emph{{Ninja: Automated Integrand Reduction via Laurent Expansion
  for One-Loop Amplitudes}},
  \href{http://dx.doi.org/10.1016/j.cpc.2014.06.017}{\emph{Comput. Phys.
  Commun.} {\bf 185} (2014) 2771--2797},
  [\href{http://arxiv.org/abs/1403.1229}{{\tt 1403.1229}}].

\bibitem{vanDeurzen:2013saa}
H.~van Deurzen, G.~Luisoni, P.~Mastrolia, E.~Mirabella, G.~Ossola and
  T.~Peraro, \emph{{Multi-leg One-loop Massive Amplitudes from Integrand
  Reduction via Laurent Expansion}},
  \href{http://dx.doi.org/10.1007/JHEP03(2014)115}{\emph{JHEP} {\bf 03} (2014)
  115}, [\href{http://arxiv.org/abs/1312.6678}{{\tt 1312.6678}}].

\bibitem{Binoth:2008uq}
T.~Binoth, J.~P. Guillet, G.~Heinrich, E.~Pilon and T.~Reiter, \emph{{Golem95:
  A Numerical program to calculate one-loop tensor integrals with up to six
  external legs}},
  \href{http://dx.doi.org/10.1016/j.cpc.2009.06.024}{\emph{Comput. Phys.
  Commun.} {\bf 180} (2009) 2317--2330},
  [\href{http://arxiv.org/abs/0810.0992}{{\tt 0810.0992}}].

\bibitem{Mastrolia:2010nb}
P.~Mastrolia, G.~Ossola, T.~Reiter and F.~Tramontano, \emph{{Scattering
  AMplitudes from Unitarity-based Reduction Algorithm at the Integrand-level}},
  \href{http://dx.doi.org/10.1007/JHEP08(2010)080}{\emph{JHEP} {\bf 08} (2010)
  080}, [\href{http://arxiv.org/abs/1006.0710}{{\tt 1006.0710}}].

\bibitem{vanHameren:2009dr}
A.~van Hameren, C.~G. Papadopoulos and R.~Pittau, \emph{{Automated one-loop
  calculations: A Proof of concept}},
  \href{http://dx.doi.org/10.1088/1126-6708/2009/09/106}{\emph{JHEP} {\bf 09}
  (2009) 106}, [\href{http://arxiv.org/abs/0903.4665}{{\tt 0903.4665}}].

\bibitem{vanHameren:2010cp}
A.~van Hameren, \emph{{OneLOop: For the evaluation of one-loop scalar
  functions}}, \href{http://dx.doi.org/10.1016/j.cpc.2011.06.011}{\emph{Comput.
  Phys. Commun.} {\bf 182} (2011) 2427--2438},
  [\href{http://arxiv.org/abs/1007.4716}{{\tt 1007.4716}}].

\bibitem{Lepage:1980jk}
G.~P. Lepage, \emph{{VEGAS}: An adaptive multidimensional integration
  program}, (1980).

\bibitem{Bern:1991aq}
Z.~Bern and D.~A. Kosower, \emph{{The Computation of loop amplitudes in gauge
  theories}}, \href{http://dx.doi.org/10.1016/0550-3213(92)90134-W}{\emph{Nucl.
  Phys.} {\bf B379} (1992) 451--561}.

\bibitem{Signer:2008va}
A.~Signer and D.~St{\"o}ckinger, \emph{{Using Dimensional Reduction for
  Hadronic Collisions}},
  \href{http://dx.doi.org/10.1016/j.nuclphysb.2008.09.016}{\emph{Nucl.Phys.}
  {\bf B808} (2009) 88--120}, [\href{http://arxiv.org/abs/0807.4424}{{\tt
  0807.4424}}].

\bibitem{Cullen:2010jv}
G.~Cullen, M.~Koch-Janusz and T.~Reiter, \emph{{Spinney: A Form Library for
  Helicity Spinors}},
  \href{http://dx.doi.org/10.1016/j.cpc.2011.06.007}{\emph{Comput. Phys.
  Commun.} {\bf 182} (2011) 2368--2387},
  [\href{http://arxiv.org/abs/1008.0803}{{\tt 1008.0803}}].

\bibitem{Jegerlehner:1985gq}
F.~Jegerlehner, \emph{{Hadronic Contributions to Electroweak Parameter Shifts:
  A Detailed Analysis}}, \href{http://dx.doi.org/10.1007/BF01552495}{\emph{Z.
  Phys.} {\bf C32} (1986) 195}.

\bibitem{Burkhardt:1989ky}
H.~Burkhardt, F.~Jegerlehner, G.~Penso and C.~Verzegnassi, \emph{{Uncertainties
  in the Hadronic Contribution to the QED Vacuum Polarization}},
  \href{http://dx.doi.org/10.1007/BF01506546}{\emph{Z. Phys.} {\bf C43} (1989)
  497--501}.

\bibitem{Jegerlehner:2006ju}
F.~Jegerlehner, \emph{{Precision measurements of $\sigma$(hadronic) for
  $\alpha$(eff)(E) at ILC energies and $(g-2)_\mu$}},
  \href{http://dx.doi.org/10.1016/j.nuclphysbps.2006.09.060}{\emph{Nucl. Phys.
  Proc. Suppl.} {\bf 162} (2006) 22--32},
  [\href{http://arxiv.org/abs/hep-ph/0608329}{{\tt hep-ph/0608329}}].

\end{thebibliography}\endgroup

\end{document}